\documentclass[pra,amsfonts,showpacs,preprint,showkeys]{revtex4}
\usepackage[T1]{fontenc}
\usepackage{graphicx}
\RequirePackage{times}
\RequirePackage{mathptm}

\begin{document}

\title{Are simultaneous Bell measurements possible?}

\author{Karl Svozil}
\email{svozil@tuwien.ac.at}
\homepage{http://tph.tuwien.ac.at/~svozil}
\affiliation{Institute for Theoretical Physics, University of Technology Vienna,
Wiedner Hauptstra\ss e 8-10/136, A-1040 Vienna, Austria}

\begin{abstract}
All experimental tests of Bell-type inequalities and Greenberger-Horne-Zeilinger setups rely on the separate and successive measurement of the terms involved.
We discuss possibilities of experimental setups to measure all relevant terms simultaneously in a single experiment and find this to be impossible.
One reason is the lack of multi-partite states which are unique in the sense that a measurement of some observable on one particle fixes the value of the corresponding observables of the other particles as well.
\end{abstract}


\maketitle
One  motivation for the beautiful Bell-type
experiment by Weihs et al. \cite{wjswz-98}
has been concerns \cite{zeilinger-86} about the implicit coincidence
between photon flight time and the specific switching frequency chosen
in one of the first experiments \cite{aspect-82b} to test Bell-type inequalities.
As pointed out by Gill et al.
\cite{Gill-Weihs-Z-Z,Gill-Weihs-Z-Z-II,gill-03}
and by Larsson and Gill  \cite{Gill-Larss-04},
any effectively computable synchronization strategy
(e.g., Ref.~\cite{pabh-04})
fails for spacelike separated observers who choose to switch the measurement directions
at random \cite{calude:94,svozil-qct}.
For Greenberger-Horne-Zeilinger (GHZ) measurements, the issue has already been discussed in the
original paper by Pan et al. \cite{panbdwz}.
Yet, there might remain reservations and uneasiness related to the fact
that in all
the experiments performed so far, different terms
in the Bell-type inequalities have been measured consecutively,
one after another, in different experiment setups.

In what follows we shall investigate, as a second and arguable conceptually more
gratifying alternative, the feasibility to either measure
or counterfactually infer
all required entities  simultaneously.
By ``simultaneous'' measurement we mean that all single measurements
are pairwise spatially separated and temporally coincide in some reference frame
(presumably in the center-of-mass frame of the particles involved).
Note also that, due to the apparent randomness and parameter independence
of the single outcomes in the correlation experiment,
relativistic locality is possible.

\begin{figure}
\begin{center}
\begin{tabular}{ccc}
\begin{tabular}{c}
\unitlength 0.80mm
\linethickness{0.4pt}
\begin{picture}(90.00,10.00)
\put(45.00,5.00){\circle{10.00}}
\put(41.00,2.00){\line(4,3){8.00}}
\put(49.00,2.00){\line(-4,3){8.00}}
\put(5.00,5.00){\oval(10.00,10.00)[l]}
\put(5.00,10.00){\line(0,-1){10.00}}
\put(2.50,5.00){\makebox(0,0)[cc]{$ \hat \alpha$}}
\put(85.00,5.00){\oval(10.00,10.00)[r]}
\put(85.00,10.00){\line(0,-1){10.00}}
\put(87.50,5.00){\makebox(0,0)[cc]{$ \hat \beta$}}
\put(40.00,5.00){\line(-1,0){33.00}}
\put(50.00,5.00){\line(1,0){33.00}}
\end{picture}
\\
$E({\hat \alpha},{\hat \beta})$
\\ \\
\unitlength 0.80mm
\linethickness{0.4pt}
\begin{picture}(90.00,10.00)
\put(45.00,5.00){\circle{10.00}}
\put(41.00,2.00){\line(4,3){8.00}}
\put(49.00,2.00){\line(-4,3){8.00}}
\put(5.00,5.00){\oval(10.00,10.00)[l]}
\put(5.00,10.00){\line(0,-1){10.00}}
\put(2.50,5.00){\makebox(0,0)[cc]{$\hat \alpha'$}}
\put(85.00,5.00){\oval(10.00,10.00)[r]}
\put(85.00,10.00){\line(0,-1){10.00}}
\put(87.50,5.00){\makebox(0,0)[cc]{$\hat \beta$}}
\put(40.00,5.00){\line(-1,0){33.00}}
\put(50.00,5.00){\line(1,0){33.00}}
\end{picture}
\\
$E({\hat \alpha'},{\hat \beta})$
\\ \\
\unitlength 0.80mm
\linethickness{0.4pt}
\begin{picture}(90.00,10.00)
\put(45.00,5.00){\circle{10.00}}
\put(41.00,2.00){\line(4,3){8.00}}
\put(49.00,2.00){\line(-4,3){8.00}}
\put(5.00,5.00){\oval(10.00,10.00)[l]}
\put(5.00,10.00){\line(0,-1){10.00}}
\put(2.50,5.00){\makebox(0,0)[cc]{$\hat \alpha$}}
\put(85.00,5.00){\oval(10.00,10.00)[r]}
\put(85.00,10.00){\line(0,-1){10.00}}
\put(87.50,5.00){\makebox(0,0)[cc]{$\hat \beta'$}}
\put(40.00,5.00){\line(-1,0){33.00}}
\put(50.00,5.00){\line(1,0){33.00}}
\end{picture}
\\
$E({\hat \alpha},{\hat \beta'})$
\\ \\
\unitlength 0.80mm
\linethickness{0.4pt}
\begin{picture}(90.00,10.00)
\put(45.00,5.00){\circle{10.00}}
\put(41.00,2.00){\line(4,3){8.00}}
\put(49.00,2.00){\line(-4,3){8.00}}
\put(5.00,5.00){\oval(10.00,10.00)[l]}
\put(5.00,10.00){\line(0,-1){10.00}}
\put(2.50,5.00){\makebox(0,0)[cc]{$\hat \alpha'$}}
\put(85.00,5.00){\oval(10.00,10.00)[r]}
\put(85.00,10.00){\line(0,-1){10.00}}
\put(87.50,5.00){\makebox(0,0)[cc]{$\hat \beta'$}}
\put(40.00,5.00){\line(-1,0){33.00}}
\put(50.00,5.00){\line(1,0){33.00}}
\end{picture}\\
$E({\hat \alpha'},{\hat \beta'})$
\end{tabular}
&  $\qquad $ &
\begin{tabular}{c}
\unitlength 0.80mm
\linethickness{0.4pt}
\begin{picture}(90.00,50.00)
\put(45.00,25.00){\circle{10.00}}
\put(41.00,22.00){\line(4,3){8.00}}
\put(49.00,22.00){\line(-4,3){8.00}}
\put(42.00,21.00){\line(-2,-1){32.00}}
\put(5.00,5.00){\oval(10.00,10.00)[l]}
\put(5.00,10.00){\line(0,-1){10.00}}
\put(2.50,5.00){\makebox(0,0)[cc]{${\hat \alpha}'$}}
\put(42.00,29.00){\line(-2,1){32.00}}
\put(5.00,45.00){\oval(10.00,10.00)[l]}
\put(5.00,40.00){\line(0,1){10.00}}
\put(2.50,45.00){\makebox(0,0)[cc]{$\hat \alpha$}}
\put(48.00,21.00){\line(2,-1){32.00}}
\put(85.00,5.00){\oval(10.00,10.00)[r]}
\put(85.00,10.00){\line(0,-1){10.00}}
\put(87.50,5.00){\makebox(0,0)[cc]{${\hat \beta}'$}}
\put(48.00,29.00){\line(2,1){32.00}}
\put(85.00,45.00){\oval(10.00,10.00)[r]}
\put(85.00,40.00){\line(0,1){10.00}}
\put(87.50,45.00){\makebox(0,0)[cc]{$\hat \beta$}}
\end{picture}
\\
$E({\hat \alpha},{\hat \beta})$ ,
$E({\hat \alpha}',{\hat \beta})$ ,
$E({\hat \alpha},{\hat \beta}')$ ,
$E({\hat \alpha}',{\hat \beta}')$
\end{tabular}
\\    \\
(a)&$\qquad $ &  (b)
\end{tabular}
\end{center}
\caption{(a) Terms entering the CHSH inequality are usually measured by four different experiments
on two-partite singlet states;
(b) Scheme for the simultaneous measurement of all expectations values for
the four directions ${\hat a}$,
${\hat a}'$,
${\hat b}$, and
${\hat b}'$
entering the CHSH inequality in a single experiment,
requiring a four-partite state with the uniqueness property.
\label{2005-hp-f1}}
\end{figure}
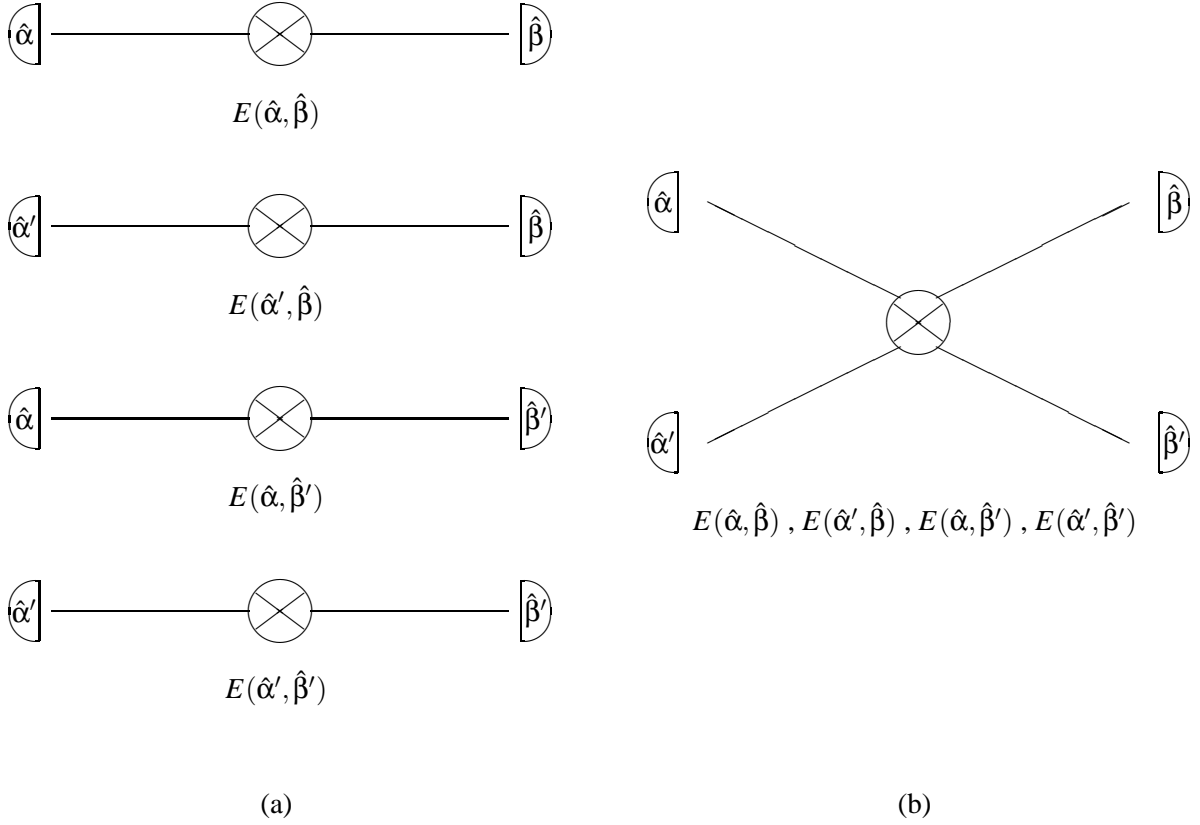

Take, for example, the Clauser-Horne-Shimony-Holt (CHSH) inequality
$
\vert
E({\hat a} ,{\hat b} )+
E({\hat a} ,{\hat b} ' )+
E({\hat a}' ,{\hat b} )-
E({\hat a} ',{\hat b} ')
\vert
\le 2
$
containing the four terms
$E({\hat a} ,{\hat b} )$,
$E({\hat a} ,{\hat b} ' )$,
$E({\hat a}' ,{\hat b} )$, and
$E({\hat a} ',{\hat b} ')$
associated with the four measurement directions
${\hat a}$,
${\hat a}'$,
${\hat b}$, and
${\hat b}'$, respectively.
Since in the usual Einstein-Podolsky-Rosen (EPR) type of setup
only two particles occur, merely two of the four directions can be measured and counterfactually
inferred in every experimental run.
As depicted in Fig.~\ref{2005-hp-f1}a),
four experimental runs with different parameter settings
${\hat a}$--${\hat b} $
${\hat a}$--${\hat b} ' $,
${\hat a}'$--${\hat b} $, and
${\hat a} '$--${\hat b} '$
are necessary to collect
the frequency counts necessary for the CHSH inequality.

In contrast to this consecutive measurements,
Fig.~\ref{2005-hp-f1}b) depicts the hypothetical configuration associated with  simultaneous
measurements all the four parameters ${\hat a}$,
${\hat a}'$,
${\hat b}$, and
${\hat b}'$ in a {\em single} experimental run;
i.e., without the necessity to change the setup for measuring different
joint probabilities.

Simultaneity presents a big conceptual challenge,
both theoretically and experimentally.
While entangled multi-partite (singlet) states abound,
the EPR argument \cite{epr} requires the unambiguous existence of (counterfactual)
{elements of physical reality.}
Theoretically, this means that we have to find four-partite states with the
{\em uniqueness property} \cite{svozil-2004-vax}
such that
knowledge of a property of one particle entails the certainty
that, if this property
were measured on the other three particles as well, the outcome of the measurement would be
a unique function of the outcome of the measurement actually performed.
More than that, such a uniqueness property must hold for all the observables
associated with the four measurement directions
${\hat a}$,
${\hat a}'$,
${\hat b}$, and
${\hat b}'$, respectively.
The uniqueness property differs from elements of physical reality
in that the latter ones,
in order to be definable by EPR type experimental setups,
require the uniqueness property.
However, absence of the uniqueness property
does not necessarily imply that no elements of physical reality exists, as
certain elements of physical reality might  not be measurable
via EPR type experimental setups.
The existence of elements of physical reality is
a necessary but not a sufficient condition for uniqueness of multi-partite quantum states.

For more than two quanta, no state satisfying the uniqueness property
has ever been proposed.
The generalized four-partite GHZ-Mermin state, for example,
has the uniqueness property only for a {\em single} measurement direction,
in which it takes a form containing only two terms
$\vert \Psi \rangle =(1/\sqrt{2})(\vert ++++\rangle + \vert ----\rangle )$;
in all other directions, $\vert \Psi \rangle$ is a superposition of $2^4$ terms,
which are by far too numerous for a uniqueness property requiring just two terms for
at least four measurement directions.
Natural candidates would be singlet states, which are form invariant with respect to changes of
the measurement directions.
Alas, group theoretical methods reveal that no singlet state of four or more particles
(in arbitrary dimensions of Hilbert space greater than one) exists which satisfies the uniqueness
property required for simultaneous measurements of more that two properties.
(States representable by only a single term in a particular measurement direction,
such as  $\vert ++++\rangle$,
have $2^4$ terms in all the other directions and are nonunique for those directions.)

Consider, for example, two-state particles; their states being labelled by
``$+$'' and ``$-$,'' respectively.
The only two-partite singlet state is the usual Bell state
$\vert \Psi_{2,2,s} \rangle =(1/\sqrt{2})(\vert + - \rangle - \vert - + \rangle )$.
There exist two four-partite singlet states (and none for three particles), namely
\begin{eqnarray}
\vert \Psi_{2,4,s_1} \rangle
&=&
{1\over 2}
\left(
\vert +- \rangle -
\vert -+ \rangle
\right)
\left(
\vert +- \rangle -
\vert -+ \rangle
\right)
,\\
\vert \Psi_{2,4,s_2} \rangle
&=&
{1\over \sqrt{3}}\left[
\vert ++-- \rangle +
\vert --++ \rangle
 \right. \nonumber \\
&&\qquad
\qquad
\left.
-  {1\over 2}
\left(
\vert +- \rangle +
\vert -+ \rangle
\right)
\left(
\vert +- \rangle +
\vert -+ \rangle
\right)
\right]
.
\label{2005-hp-ep24s2}
\end{eqnarray}
Both of these four-partite states are nonunique in all directions,
for detection of, say, state ``$-$'' on the first particle leaves open the possibility
to find the third and fourth particles either in states ``$+$'' or in ``$-$''
for $\vert \Psi_{2,4,s_1} \rangle$.
Likewise,
the proposition
\begin{quote}
{\em `The first particle has spin state ``$-$''.'}
\end{quote}
does not fix a single term  of $\vert \Psi_{2,4,s_2} \rangle$
in Eq.~(\ref{2005-hp-ep24s2}), but rather leaves open one of the two possibilities
\begin{quote}
{\em `The second (third, fourth) particle has spin state ``$-$'';'}\\
{\em `The second (third, fourth) particle has spin state ``$+$'';'}
\end{quote}
both occurring at random.

This ambiguity gets worse as the number of particles increases.
The $2n$-partite singlet states (singlet states with an odd number of particles do not exist)
with the least number of terms are
\begin{equation}
\vert \Psi_{2,2n,s_1} \rangle
= {1\over 2^{n/2}}\left(
\vert +- \rangle -
\vert -+ \rangle
\right)^n .
\end{equation}
For $n>1$, they are nonunique.

Also for singlet states of three-state particles, the uniqueness limit is reached for two particles;
i.e., for (``$\pm$'' means here ``$\pm 1$'') $
\vert \Psi_{3,2,s} \rangle
= ({1/ \sqrt{3}})(
\vert + -\rangle
+
\vert - +\rangle
-
\vert 0 0\rangle
)$.
The only singlet state of three spin one particles is
\begin{equation}
\vert \Psi_{3,3,s} \rangle
= {1\over \sqrt{6}}\left(
\vert - + 0\rangle
-
\vert - 0 +\rangle
+
\vert + 0 - \rangle
-
\vert + - 0\rangle
+
\vert 0 - + \rangle
-
\vert 0 + - \rangle
\right),
\end{equation}
which does not allow unambiguous counterfactual reasoning.
The situation gets worse for singlet states of four or more spin one quanta,
as all of the three singlet states $\vert \Psi_{3,4,s_1}\rangle , \vert \Psi_{3,4,s_2}\rangle , \vert \Psi_{3,4,s_3}\rangle$
of four spin one quanta
\begin{eqnarray}
\vert \Psi_{3,4,s_1}\rangle  &=&
\frac{1}{\sqrt{5}}
\left[
\frac{2}{3}
\vert0000\rangle
+
\vert --++\rangle  +\vert ++--\rangle   \right.
\nonumber \\
&&\quad
- \frac{1}{2}
\left(
\vert-00+\rangle
+\vert0-0+\rangle
+\vert-0+0\rangle
+\vert0-+0\rangle \nonumber  \right. \\
&& \left. \quad
\qquad
+\vert0+-0\rangle
+\vert +0-0\rangle
+\vert0+0-\rangle
+\vert+00-\rangle
\right)\nonumber \\
&&\quad
+\frac{1}{3}
\left(
\vert00-+\rangle
+\vert -+00\rangle
+\vert+-00\rangle
+\vert00+-\rangle
\right)\nonumber \\
&&\quad
+
\left. \frac{1}{6}
\left(
\vert-+-+\rangle
+\vert+--+\rangle
+\vert-++-\rangle
+\vert+-+-\rangle
\right)
\right] ,
 \\
\vert \Psi_{3,4,s_2}\rangle  &=&\frac{1}{2{\sqrt{3}}}
\left(
\vert -00+\rangle
- \vert 0-0+\rangle
- \vert 0+0-\rangle
+\vert +00-\rangle \right. \nonumber \\
&&\quad
- \vert -0+0\rangle
+\vert 0-+0\rangle
+\vert 0+-0\rangle
- \vert +0-0\rangle \nonumber    \\
&&\quad \left.
+\vert -++-\rangle
- \vert +-+-\rangle
- \vert -+-+\rangle
+\vert +--+\rangle
\right) ,
  \\
  \vert \Psi_{3,2n,s_3}\rangle &=& \frac{1}{3^{n/2}}
\left(
\vert + -\rangle
+
\vert - +\rangle
-
\vert 0 0\rangle
\right)^n ,
\end{eqnarray}
are nonunique.
From all  singlet states of $2n$ spin-one particles,
$\vert \Psi_{3,2n,s_3}\rangle$ contains the least number of terms.
For $n>1$, they are nonunique.
In general, for an even number of $2n$ spin-$j$ particles,
the singlet states containing the least number of terms
can be written as
$
\vert \Psi_{j,2n,s}\rangle = \left(\vert \Psi_{j,2,s}\rangle \right)^n
$,
where $\vert \Psi_{j,2,s}\rangle$ is the two-partite singlet state.

Classically simultaneous measurements can be performed
on an arbitrary number of particles and for arbitrary many observables.
They are not bound by the quantum no-cloning theorem and complementarity.

Indeed, quantum probabilities do not allow the consistent co-existence of
classical truth tables already for the usual Bell-type configurations such as CHSH
\cite{peres222,svozil-krenn}.
We therefore conclude that it is impossible to construct quantum states
of four or more particles with the uniqueness
property for four or more directions.
Likewise, because of nonuniqueness,
the observables involved in a Kochen-Specker-type argument
cannot be measured simultaneously.

Let us briely summarize as follows.
The proposed scheme is not different from the standard treatment of EPR-type setups,
since elements of physical reality are not directly measured
but counterfactually inferred.
One advantage of simultaneous measurements and inference might be  the
direct evaluation of all quantities involved; as compared to
a serial, consecutive approach which is usually adopted.
However, as the EPR argument is based on the uniqueness of the counterfactual properties,
any such method fails already for more than two entangled particles.

A new kind of protection scheme seems to emerge which,
due to the impossibility to find states with the required uniqueness property,
forbids more direct experiments
to measure all terms in Kochen-Specker-type,
Bell-type or GHZ-type configurations at once.
It is straightforward to deduce nonuniqueness
for singlet states of three or more particles.
An indirect argument, based on Bell- and Kochen-Specker-type theorems,
proves the nonexistence of general states with the uniqueness property.


\end{document}